\documentclass[12pt]{article}
\usepackage{epsfig,amsfonts,amssymb}
\usepackage{amsmath,graphics}
\usepackage{slashed}
\usepackage{color}
\usepackage{xcolor}

 \usepackage{hyperref}
\usepackage{cleveref}
\crefformat{footnote}{#2\footnotemark[#1]#3}
\def\ZZZ{{\hbox{ Z\kern-1.6mm Z}}}
\def\RRR{{\hbox{ R\kern-2.4mm R}}}
\def\CCC{{\hbox{ C\kern-2.0mm C}}}
\def\zzz{{\hbox{z\kern-1mm z}}}

\def\one{{\hbox{ 1\kern-.8mm l}}}
\def\zero{{\hbox{ 0\kern-1.5mm 0}}}

\newcommand{\be}{\begin{equation}}
\newcommand{\bea}{\begin{eqnarray}}
\newcommand{\ee}{\end{equation}}
\newcommand{\eea}{\end{eqnarray}}

\begin{document}

\baselineskip 24pt

\begin{center}
{\Large \bf On the Gravitini Zero Modes riding on top of \footnote{We have borrowed the phrase ``riding on top of'' from the first sentence of the abstract of  \cite{Brooks:1995wb}, where it was used in the context of fermion zero modes.} \\ Multiple Black Holes}

\end{center}

\vskip .6cm
\medskip

\vspace*{4.0ex}

\baselineskip=18pt

\centerline{\large \rm   Palash Dubey$^{1}$, Chethan N. Gowdigere$^{2}$}

\vspace*{4.0ex}

\centerline{\large \it National Institute of Science Education and Research Bhubaneshwar,}

\centerline{\large \it  P.O. Jatni, Khurdha, 752050, Odisha, INDIA}

\vspace*{1.0ex}

\centerline{\large \it Homi Bhabha National Institute, Training School Complex, }

\centerline{\large \it  Anushakti Nagar, Mumbai 400094, INDIA}

\vspace*{4.0ex}
\centerline{E-mail: $^1$palash@niser.ac.in, $^2$chethan.gowdigere@niser.ac.in}

\vspace*{5.0ex}

\centerline{\bf Abstract} \bigskip

We study gravitini zero modes in four dimensional $\mathcal{N} = 2$ pure supergravity theory. The 
gravitini zero modes we study are solutions to the Rarita-Schwinger equations of motion in the background of the purely bosonic Majumdar-Papapetrou background. We start with a very generic ansatz for the gravitini that involves $32$ ansatz functions and reduce the Rarita-Schwinger equations to a set of linear coupled partial differential equations on $\mathbf{R}^3$: the kind familiar from electromagnetism with divergences, curls etc. We first show how the gravitini zero modes due to broken supersymmetries arise in this set-up and how they solve the equations on $\mathbf{R}^3$. Then we go on and obtain other solutions to these equations: the `extra' fermion zero modes. If we count the fermion zero mode due to broken supersymmetries as one spinor worth of solutions, we have obtained three other spinors worth of solutions.

\vfill \eject

\baselineskip=18pt

\tableofcontents

\section{Introduction}

In this paper we study fermion zero modes riding on top of purely bosonic supersymmetric solutions  in four dimensional supergravity. The supergravity theory we study is the $\mathcal{N} = 2$ pure supergravity \cite{Ferrara:1976fu}, which is a theory of a graviton, a graviphoton and a two gravitinis. The purely bosonic solution we focus on is the multi-centre Majumdar-Papatetrou \cite{Majumdar:1947eu},\cite{Papapetrou:1948jw} black hole solution, which is known to be supersymmetric  \cite{Gibbons:1982fy}, \cite{Hull:1983ap}, \cite{Tod:1983pm} preserving half and breaking the other half of the supersymmetries of the action.  

An important properly of purely bosonic supersymmetric solutions is that they admit fermion zero modes which are classical solutions to the fermionic equations of motion, which in this case are the Rarita-Schwinger equations, in the background of the bosonic solution. It is straight forward to generate some of these fermion zero modes, the ones corresponding to the super symmetries broken by the bosonic solution (the goldstinos).  One only has to perform a supersymmetry transformation of the bosonic solution using a supersymmetry parameter, that at asymptotic infinity is orthogonal to the Killing spinors of the bosonic solution; these supersymmetry parameters are referred to as anti-Killing spinors. The resulting configuration of fields will be unchanged for the bosonic fields but now non-zero for the fermionic fields. This non-zero fermionic field thus obtained is guaranteed to solve the fermionic equations of motion in the bosonic background.  We do work out the explicit form of these fermion zero modes due to broken supersymmetry here in this paper. This is not new work; in fact it (and more) was done in a series of papers by Aichelburg et al \cite{Aichelburg:1983ux}, \cite{Aichelburg:1986wv}, \cite{Aichelburg:1987hy}, \cite{Aichelburg:1987hx}, \cite{Aichelburg:1987hz}, \cite{Aichelburg:1987ia} and moreover they have computed the full fermionic solution, not just up to the linearised order, which is what we restrict ourselves to in this paper. 

The new work in this paper is providing a computational set-up for and going on and explicitly obtaining some  fermion zero modes in the background of the purely bosonic Majumdar-Papapetrou supersymmetric solution. We come up with an ansatz for the fermions and directly solve the fermion equations of motion. We find the fermion zero mode due to broken supersymmetries as one of the solutions to the fermion equations of motion as well as other fermion zero modes.  If we count the fermion zero mode due to broken supersymmetries as one spinor worth of solutions, we would have obtained three other spinor worth of solutions. These ``extra fermion zero modes'' are the main result of this paper. 

Most of the papers on this subject \cite{Aichelburg:1983ux}-\cite{Brooks:1995wb}  were written decades ago.  The motivations for those papers (should and partially) does serve as motivation for our work. That being the study of supersymmetric solitons in supergravity theories and more particularly finding the supersymmetric theory that describes the dynamics of these solitons. For the Majumdar-Papapetrou  black hole solitons, the theory was found in \cite{Brooks:1995wb}.

We are also driven by a more recent motivation. This comes from the microscopic theory for the entropy of multiple supersymmetric black holes. In certain cases, considerations coming from microscopic formulae, indicate that the macroscopic bosonic supersymmetric black hole solution should admit extra fermion zero mode solutions, beyond the fermion zero modes due to broken super symmetries.  This is known to occur for both  theories and  solutions different from the ones being studied here in this paper. Extra fermion zero modes are expected in multi-centre Denef-Bates \cite{Bates:2003vx} solutions which are solutions that preserve (generically) one quarter of the supersymmetries of four dimensional $\mathcal{N} = 4$ supergravity. Some references for extra fermion zero modes are \cite{Banerjee:2008pv}\cite{Dabholkar:2008zy}\cite{Dabholkar:2010rm} \cite{Bossard:2019ajg}.
 Both the solutions and the theories are very involved.  With that distant goal in mind, we start off in this paper with studying fermion zero modes in supersymmetric multi-centre black holes in the simplest possible setting in which they exist.  That is, for the supersymmetric Majumdar-Papapetrou multi-black holes in four dimensional $\mathcal{N}=2$ pure supergravity.  We hope \cite{work} the set-up and the results of this paper goes some way in getting to the above mentioned goal. 

This paper is organised as follows. The section (second) that immediately follows the present one reviews relevant details of the supergravity, the supersymmetry transformations, the Majumdar-Papapetrou solutions, their Killing spinors, computes the anti-Killing spinors and with those for supersymmetry parameters generates the fermion zero modes due to broken supersymmetries. In the third section, first we motivate the ansatz for the fermions, which turns out to have a total of $32$ real functions and then proceed to compute the consequence of the fermion equations of motion on these ansatz functions. It turns out that the $32$ functions form two decoupled sectors of $16$ functions each; the functions in each sector  are coupled among themselves in a set of coupled first order linear partial differential equations. We give the equations for both the sectors in this section (third) and we also discuss the $\gamma$-tracelessness gauge condition that we work with through out the paper.  Then, in section four, we proceed to solve the equations of motion in one of the decoupled sectors, the one that includes the ansatz functions that are turned on in the fermion zero mode due to broken supersymmetry. Before solving, we show that the fermion zero mode due to broken supersymmetries indeed solves the equation of motion. Then we present the analysis of the equations that leads to the extra fermion zero modes. In the last section (five) we conclude with a discussion and directions for future work.

\section{Gravitini Zero Modes due to Broken Supersymmetries}

In this section, we will describe the relevant details needed, on the supergravity theory, the supersymmetric solutions, the Killing spinors and the fermion zero modes due to broken supersymmetry. 

\subsection{Four dimensional $\mathcal{N} = 2$ pure supergravity}

The fundamental fields of four dimensional $\mathcal{N} = 2$ pure supergravity are (i) the gravitational field written in terms of the vierbien fields $e^a = e^a_{\mu} \, dx^\mu$, (ii) the electromagnetic potential one-form $A = A_{\mu}\,dx^\mu$ and (iii) two Majorana spinor-valued one-forms $\psi^j, ~j = 1,2$ combined to a complex Dirac field $\psi = \psi^1 + i \, \psi^2 = \psi_\mu\,dx^\mu$. All fields are one-forms: the bosonic fields are ordinary one-forms i.e. taking values in space-time functions and the fermionic field is a one-form taking values in Dirac spinors. One can thus suppress the indices corresponding to one-form components.

The two supersymmetry transformation parameters $\epsilon^j, ~j=1,2$ are Majorana spinor fields, combined into a complex Dirac spinor field $\epsilon = \epsilon^1 + i \,\epsilon^2$. The supersymmetry transformations of the theory are :
\bea
\label{1.0}
\delta e^a = - \frac{ik}{2} \left[ \overline{\epsilon}\,\gamma^a\,\psi - \overline{\psi}\,\gamma^a\,\epsilon\right], \qquad  \delta A  = \frac{i}{2} \left[ \overline{\epsilon}\,\psi - \overline{\psi}\,\epsilon\right], \qquad  \delta \psi  = \frac{1}{k} \, \hat{\mathcal{D}} \epsilon.
\eea
where 
\begin{equation}
\label{2.0}
\hat{\mathcal{D}} = d + \frac12 \omega^{ab}\sigma_{ab} - \frac{k}{2} \hat{F}^{ab} \sigma_{ab} \gamma
\end{equation}
with 
\be
\hat{F} = F - \frac{ik}{2}\,\overline{\psi} \wedge \psi, \qquad \gamma \equiv \gamma_a e^a.
\ee
$\omega^{ab}$ is the spin-connection one-form and $F$ is the electromagnetic field strength two-form and $\gamma$ is a Clifford-algebra valued one-form built out of the vierbien one-forms. We work with the conventions of \cite{Aichelburg:1986wv}, \cite{Embacher:1985ie}, \cite{Aichelburg:1980up}, \cite{Brooks:1995wb}. The space-time signature is mostly negative and $k^2 = 4\pi G$ and we use a Weyl representation for the $\gamma$-matrices \footnote{\begin{eqnarray} \label{4.0}
\gamma_0 = \begin{pmatrix} 0& 1 \\ 1& 0 \end{pmatrix},  \gamma_1 = \begin{pmatrix} 0& \sigma_1 \\ -\sigma_1 & 0 \end{pmatrix},  \gamma_2 = \begin{pmatrix} 0& \sigma_2 \\ -\sigma_2 & 0 \end{pmatrix}, \quad \gamma_3 = \begin{pmatrix} 0& \sigma_3 \\ - \sigma_3 & 0 \end{pmatrix}, 
\sigma_{ab} = \frac14 [ \sigma_a, \sigma_b ], \gamma_5 = i \gamma_0\gamma_1\gamma_2\gamma_3 \end{eqnarray}}.

The action of four dimensional $\mathcal{N}=2$ supergravity, which can be found in \cite{Ferrara:1976fu}, which we won't write here, is invariant under the supersymmetry transformations \eqref{1.0}. Amongst the equations of motion, we will need explicitly only the linearised fermionic equations of motion:
\be
\label{5.0}
\gamma \wedge \hat{\mathcal{D}} \wedge \psi = 0
\ee
The $\hat{\mathcal{D}}$ that is in \eqref{5.0} is the linearised version of the one in \eqref{2.0}. This same linearised version is also relevant for the action of supersymmetry on a purely bosonic configuration and is given explicitly below. In the rest of the paper $\hat{\mathcal{D}}$ will denote this linearised version.

\bea
\label{6.0}
\hat{\mathcal{D}} &&= d + \frac12 \omega^{ab}\sigma_{ab} - \frac{k}{2} F^{ab} \sigma_{ab} \gamma \equiv d + \omega F 
\eea
$\hat{\mathcal{D}}$ is an operator taking values in the Clifford-algebra: the first term $d$ has no $\gamma$ matrices, the second term $\frac12 \omega^{ab}\sigma_{ab}$ has two $\gamma$ matrices and the third term $- \frac{k}{2} F^{ab} \sigma_{ab} \gamma$ has one or three $\gamma$ matrices. The second and third terms of $\hat{\mathcal{D}}$, denoted by  $\omega F$, is a Clifford-algebra valued one-form. The action of $\hat{\mathcal{D}}$ is on a spinor (which can be thought of as a spinor valued zero form) to produce a spinor valued one-form: $d$ acts on spinor valued zero-forms and produces spinor valued one-forms while $\omega F$ being Clifford-algebra valued acts on spinor to produce a spinor and since it is also a one-form it produces a spinor-valued one-form in the end.

\subsection{Majumdar-Papapetrou multi-centre black holes}

The Majumdar-Papapetrou (MP) solution\cite{Majumdar:1947eu},\cite{Papapetrou:1948jw} is a purely bosonic solution to four dimensional $\mathcal{N} = 2$ pure supergravity describing multiple extremal static black holes. The non-vanishing fields on the MP solution are: 
\be
\label{8.0}
e^0 = \frac{dt}{V}, \quad e^1 = V\, dx, \quad e^2 = V\, dy, \quad e^3 = V\, dz  \quad A = -\frac{dt}{k\,V}.
\ee
For the following choice of the function $V$, 
\be
V(x,y,z) = 1 + \sum_{J=1}^n \frac{G M_{J}}{|\vec{x}-\vec{x}_{J}|}, \quad \vec{x}_J \neq \vec{x}_K, \quad J \neq K
\ee
this bosonic configuration \eqref{8.0} describes $n$ static extremal charged black holes with mass parameters $M_J$ and electric charges $k M_J$. Without loss of generality, we restrict ourselves to the 
case of positive electric charges and zero magnetic charges. The sign of all electric charges may be reversed by changing the sign of $A$, while magnetic charge can be generated by a duality transformation. The spatial part of the metric is conformal to a $\mathbf{R}^3$ with co-ordinates $x,y,z$ and this $\mathbf{R^3}$ will play a crucial role for us, 

\subsubsection{The $\hat{\mathcal{D}}$ operator for the Majumdar-Papapetrou solution}

There is a $\hat{\mathcal{D}}$ operator \eqref{6.0} for every bosonic background.  We first construct the $\hat{\mathcal{D}}$ operator for the MP solution. This is relevant to compute the Killing spinors of the MP solution as well as to compute the fermion zero modes in the MP solution.  
\begin{eqnarray} \label{10.0}
\hat{\mathcal{D}} = d &+& \omega F \nonumber \\
= d &+& e^0 \left[ \frac{\partial_x V}{2V^2}\,\gamma_1 (\mathbf{1}- \gamma_0) + \frac{\partial_y V}{2V^2}\,\gamma_2 (\mathbf{1}- \gamma_0) + \frac{\partial_z V}{2V^2}\,\gamma_3 (\mathbf{1}- \gamma_0) \right] \nonumber \\  
&+& e^1 \left[ \frac{\partial_x V}{2V^2}\,\gamma_0 - \frac{\partial_y V}{2V^2}\,\gamma_1 \gamma_2 (\mathbf{1}- \gamma_0) + \frac{\partial_z V}{2V^2}\,\gamma_3 \gamma_1 (\mathbf{1}- \gamma_0) \right] \nonumber \\
&+& e^2 \left[ \frac{\partial_y V}{2V^2}\,\gamma_0 - \frac{\partial_z V}{2V^2}\,\gamma_2 \gamma_3 (\mathbf{1}- \gamma_0) + \frac{\partial_x V}{2V^2}\,\gamma_1 \gamma_2 (\mathbf{1}- \gamma_0) \right] \nonumber \\
&+& e^3 \left[ \frac{\partial_z V}{2V^2}\,\gamma_0 - \frac{\partial_x V}{2V^2}\,\gamma_3 \gamma_1 (\mathbf{1}- \gamma_0) + \frac{\partial_y V}{2V^2}\,\gamma_2 \gamma_3 (\mathbf{1}- \gamma_0) \right].
\end{eqnarray}

Killing spinor fields, $\epsilon_{KS}$ are those, when used to perform a supersymmetry transformation on a configuration of fields, leaves it invariant and henceforth termed supersymmetric. The MP solution is a supersymmetric solution to $4d$ $\mathcal{N}=2$ supergravity; this was shown in \cite{Gibbons:1982fy}, \cite{Hull:1983ap}, \cite{Tod:1983pm}. For a purely bosonic configuration of fields, such as the MP solution, the Killing spinors solve the following vanishing-fermion-variation equation
\begin{equation}
\label{11.0}
\hat{\mathcal{D}} \epsilon_{KS} = 0.
\end{equation}
with the bosonic-variation-equations trivially satisfied due to the vanishing of the fermion fields in the purely bosonic configuration. 
From equation \eqref{10.0}, we have 
\begin{eqnarray} \label{12.0}
0= &e^0& \left[ V \partial_t\epsilon_{KS}  + \frac{\partial_x V}{2V^2}\,\gamma_1 (\mathbf{1}- \gamma_0) \epsilon_{KS} + \frac{\partial_y V}{2V^2}\,\gamma_2 (\mathbf{1}- \gamma_0)\epsilon_{KS}+ \frac{\partial_z V}{2V^2}\,\gamma_3 (\mathbf{1}- \gamma_0)\epsilon_{KS} \right] \nonumber \\  
+ &e^1& \left[ \frac{\partial_x \epsilon_{KS}}{V}  + \frac{\partial_x V}{2V^2}\,\gamma_0 \epsilon_{KS} - \frac{\partial_y V}{2V^2}\,\gamma_1 \gamma_2 (\mathbf{1}- \gamma_0) \epsilon_{KS} + \frac{\partial_z V}{2V^2}\,\gamma_3 \gamma_1 (\mathbf{1}- \gamma_0) \epsilon_{KS} \right] \nonumber \\
+ &e^2& \left[ \frac{\partial_y \epsilon_{KS}}{V}  +\frac{\partial_y V}{2V^2}\,\gamma_0 \epsilon_{KS} - \frac{\partial_z V}{2V^2}\,\gamma_2 \gamma_3 (\mathbf{1}- \gamma_0) \epsilon_{KS} + \frac{\partial_x V}{2V^2}\,\gamma_1 \gamma_2 (\mathbf{1}- \gamma_0) \epsilon_{KS} \right] \nonumber \\
+ &e^3& \left[ \frac{\partial_z \epsilon_{KS}}{V}  + \frac{\partial_z V}{2V^2}\,\gamma_0 \epsilon_{KS} - \frac{\partial_x V}{2V^2}\,\gamma_3 \gamma_1 (\mathbf{1}- \gamma_0) \epsilon_{KS}+ \frac{\partial_y V}{2V^2}\,\gamma_2 \gamma_3 (\mathbf{1}- \gamma_0) \epsilon_{KS} \right].
\end{eqnarray}
The (spatial) Clifford algebra is spanned by $\mathbf{1}, \gamma_1, \gamma_2, \gamma_3, \gamma_1 \gamma_2, \gamma_2 \gamma_3, \gamma_3 \gamma_1$ and $\gamma_1 \gamma_2 \gamma_3$; the one forms are spanned by $e^0, e^1, e^2$ and $e^3$. Hence the co-efficient of every Clifford-algebra basis element in each line of \eqref{12.0} should separately vanish. 
From the first line of \eqref{12.0}, from the basis element $\mathbf{1}$ we get the equation
\begin{equation} \label{13.0}
\partial_t \epsilon_{KS} = 0
\end{equation}
and from each of the basis elements $\gamma_1, \gamma_2$ and $\gamma_3$ we get the equation 
\begin{equation}\label{14.0}
\gamma_0 \epsilon_{KS} = \epsilon_{KS}
\end{equation}
Equation \eqref{13.0}  is the expectation that the Killing spinors supported by a time-independent background are time-independent. Equation \eqref{14.0}, using the form of $\gamma_0$ from \eqref{4.0}, implies that 
\begin{equation}
\epsilon_{KS} \sim \begin{pmatrix} c  \\ c \end{pmatrix}
\end{equation}
where $c$ is a constant two-component spinor. At this stage, $\epsilon_{KS}$ can be multiplied by a overall function of space.  From the second line of \eqref{12.0}, the co-efficients of the basis elements $\gamma_1\gamma_2, \gamma_3\gamma_1$ vanish due to \eqref{14.0} and the equation from the basis element $\mathbf{1}$ fixes the $x$-dependence of the overall factor. Similarly, the $y$ and $z$-dependence are fixed from the third and fourth lines of \eqref{12.0} and we obtain
\begin{equation} \label{16.0}
\epsilon_{KS} = \frac{1}{\sqrt{V}} \begin{pmatrix} c  \\ c \end{pmatrix}.
\end{equation}

\subsubsection{The anti-Killing spinors of the Majumdar-Papapetrou solution}
Anti-Killing spinors\footnote{\label{noter}This term is again borrowed from \cite{Brooks:1995wb}.}, $\epsilon_{AKS}$  are spinor fields which correspond to broken supersymmetries and hence at asymptotic infinity should converge to spinors orthogonal to those of Killing spinor fields. That is $\epsilon_{AKS}$ is  proportional to $\begin{pmatrix} c  \\ -c \end{pmatrix}$ which means 
\begin{equation}\label{17.0}
\gamma_0 \epsilon_{AKS} = -\epsilon_{AKS}, \qquad (\mathbf{1}- \gamma_0) \epsilon_{AKS} = 2\, \epsilon_{AKS}.
\end{equation}
At this stage we have 
\begin{equation}
\epsilon_{AKS} \sim \begin{pmatrix} c  \\ -c \end{pmatrix} \nonumber 
\end{equation}
and we can multiply with any function of space-time. We will choose this factor so that the resulting gravitino satisfies the $\gamma$-traceless gauge condition, which is the gauge condition employed throughout this paper.  That is 
\begin{eqnarray} 
0 = && \gamma^\mu \hat{\mathcal{D}}_\mu\,\epsilon_{AKS} \nonumber \\
0 = &&- V \partial_t\epsilon_{AKS}   - \gamma_1 \frac{\partial_x \epsilon_{AKS}}{V} -  \frac{\partial_x V}{2V^2}\,\gamma_1  \epsilon_{AKS} - \gamma_2 \frac{\partial_y \epsilon_{AKS}}{V} - \frac{\partial_y V}{2V^2}\,\gamma_2 \epsilon_{AKS} - \gamma_3 \frac{\partial_z \epsilon_{AKS}}{V}  -  \frac{\partial_z V}{2V^2}\,\gamma_3 \epsilon_{AKS}  \nonumber 
\eea
which says the function is time-independent and we obtain 
\begin{equation}\label{18.0}
\epsilon_{AKS} = \frac{1}{\sqrt{V}} \begin{pmatrix} c  \\ - c \end{pmatrix}.
\end{equation}

\subsubsection{Gravitini zero modes due to broken supersymmetries}
One can generate fermion zero modes, i.e. solutions to the fermion equations of motion in the purely bosonic supersymmetric background, by performing a supersymmetry transformation whose supersymmetry parameters are the anti-Killing spinors; note that the anti-Killing spinors do not vanish at infinity and hence they provide for a non-trivial supergauge transformation.  From \eqref{1.0}, we obtain
\begin{equation}
\psi_{bs} = \frac{1}{k} \hat{\mathcal{D}}\, \epsilon_{AKS},
\end{equation}
where $bs$ stands for ``broken supersymmetries.'' Using the explicit form for the $\hat{\mathcal{D}}$ operator for the MP background in \eqref{10.0} we have 
\begin{eqnarray}\label{20.0}
k\,\psi_{bs} =  &e^0& \left[ \frac{\partial_x V}{V^2}\,\gamma_1 + \frac{\partial_y V}{V^2}\,\gamma_2  + \frac{\partial_z V}{V^2}\,\gamma_3 \right] \epsilon_{AKS} \nonumber \\  
+ &e^1& \left[ -\frac{\partial_x V}{V^2}  - \frac{\partial_y V}{V^2}\,\gamma_1 \gamma_2  + \frac{\partial_z V}{V^2}\,\gamma_3 \gamma_1  \right] \epsilon_{AKS} \nonumber \\
+ &e^2& \left[- \frac{\partial_y V}{V^2}\, - \frac{\partial_z V}{V^2}\,\gamma_2 \gamma_3   + \frac{\partial_x V}{V^2}\,\gamma_1 \gamma_2   \right] \epsilon_{AKS} \nonumber \\
+ &e^3& \left[- \frac{\partial_z V}{V^2}\, - \frac{\partial_x V}{V^2}\,\gamma_3 \gamma_1  + \frac{\partial_y V}{V^2}\,\gamma_2 \gamma_3   \right] \epsilon_{AKS}.\end{eqnarray}
From the above equation, we see that the  gravitini field, for the fermion zero mode associated with broken supersymmetries, is a  spatial Clifford-algebra element multiplying a constant spinor and also a one-form.  This motivates the ansatz in the next section.

\section{The Equations for Gravitini Zero Modes}

In this section, we will present the generic ansatz for the fermions and set-up the fermion equations of motion for this ansatz. We will also present the gauge condition and impose the gauge conditions on the equations of motion to arrive at the final set of equations which need to be solved. 

\subsection{An ansatz for the Gravitini}
The four dimensional gravitino field $\psi$ consists of four Dirac spinors and as such has $4 \times 8$ real functions, where the $8$ is the number of real functions associated to a Dirac spinor field.  We will think of the gravitino field as taking values in the Clifford algebra and acting on the constant spinor $\begin{pmatrix} c  \\ - c \end{pmatrix}$, generalising the fermion zero mode due to broken supersymmetry \eqref{20.0}. In our ansatz the 32 real functions associated to $\psi$ will be organised again as $4 \times 8$ but now $8$ is the number of real functions associated to a  spatial Clifford algebra field:
\begin{eqnarray} \label{21.0}
\psi_{gen} =  &e^0& \left[ {{h_0}} \mathbf{1} + h_1\,\gamma_1 + h_2\,\gamma_2  + h_3\,\gamma_3 + {{h_{12}}} \gamma_1\gamma_2 + {{h_{23}}} \gamma_2\gamma_3 + {{h_{31}}} \gamma_3\gamma_1 + {{h_{123}}} \gamma_1\gamma_2\gamma_3 \right] \epsilon_0 \nonumber \\  
+ &e^1& \left[ f_1\mathbf{1} + {{f_{11}}}\,\gamma_1 + {{f_{12}}}\,\gamma_2  + {{f_{13}}}\,\gamma_3  + f_{112}\,\gamma_1 \gamma_2  + {{f_{123}}}\,\gamma_2 \gamma_3  - f_{113}\,\gamma_3 \gamma_1  + {{f_{1123}}}  \gamma_1\gamma_2\gamma_3 \right] \epsilon_0 \nonumber \\
+ &e^2& \left[ f_2\mathbf{1} + {{f_{21}}}\,\gamma_1 + {{f_{22}}}\,\gamma_2  + {{f_{23}}}\,\gamma_3 - f_{221}\,\gamma_1 \gamma_2 + f_{223}\,\gamma_2 \gamma_3  + {{f_{231}}}\,\gamma_3 \gamma_1 +  {{f_{2123}}} \gamma_1\gamma_2\gamma_3  \right] \epsilon_0 \nonumber \\
+ &e^3& \left[ f_3\mathbf{1}  + {{f_{31}}}\,\gamma_1 + {{f_{32}}}\,\gamma_2  + {{f_{33}}}\,\gamma_3 + {{f_{312}}}\,\gamma_1 \gamma_2  - f_{332}\,\gamma_2 \gamma_3 + f_{331}\,\gamma_3 \gamma_1  +  {{f_{3123}}} \gamma_1\gamma_2\gamma_3 \right]  \epsilon_0. \nonumber \\
\end{eqnarray}
where $\epsilon_0 = \begin{pmatrix} c  \\ - c \end{pmatrix}$ is the constant spinor that the Anti-Killing spinor field asymptotes to at spatial infinity. It will be useful to defined the following vectors in $\mathbf{R}^3$:
\bea 
\label{22.0}
\vec{h} = (h_1, h_2, h_3), \quad \vec{f} = (f_1, f_2, f_3), \quad \vec{f_B} = (f_{221}, f_{332}, f_{113})\quad \vec{f_C} = (f_{331}, f_{112}, f_{223}), \nonumber \\ 
{{\vec{h}_B = (h_{23}, h_{31}, h_{12})}}, \quad {{\vec{f_D} = (f_{23}-f_{32}, f_{31}-f_{13}, f_{12}-f_{21})}}, \quad  {{\vec{f_E} = (f_{1123}, f_{2123}, f_{3123})}}.
\eea

It is useful to record, what values the  ansatz functions  take in $\psi_{bs}$. Comparing \eqref{20.0} and \eqref{21.0}, we see
\bea
\label{23.0}
\vec{h} = -\vec{f} = - \vec{f_B} = - \vec{f_C} =  \frac{1}{k}\,\frac{\vec{\nabla}V}{V^{\frac52}} \qquad \text{for} \qquad\psi_{bs}
\eea
and all other ansatz functions vanish.  The vector field $\frac{\vec{\nabla}V}{V^{\frac52}}$ has the right boundary conditions both at asymptotic infinity, where it vanishes as $\mathcal{O}(r^{-2})$, and is regular at each of the horizons, where it vanishes as $\mathcal{O}(r^{\frac12})$; $r = 0$ is the location of the black hole horizon. 

\subsection{Rarita-Schwinger equation in the MP background}
 We plug in the ansatz \eqref{21.0} into the Rarita-Schwinger equation in the MP background:
 \be
 \label{23.9}
 \gamma\,  \wedge d\psi + \gamma \, \wedge \, \omega F \, \wedge \psi = 0
 \ee
where $\omega F$ is defined in \eqref{10.0}. Each of the terms in the left hand side above is a three-form taking values in the spatial Clifford algebra. For any such Clifford-algebra valued three-form $\mathcal{E}$, introduce the following notation:
\begin{eqnarray}\label{cl}
\mathcal{E} &=& e^0 \wedge e^1 \wedge e^2\, \left[ \, \left(\mathcal{E}\right)_{012}^{\mathbf{1}}\,\mathbf{1} + \left(\mathcal{E}\right)_{012}^{\mathbf{\gamma_1}}\,\gamma_1 +  \ldots \right] \nonumber \\
&+& e^0 \wedge e^2 \wedge e^3\, \left[ \, \left(\mathcal{E}\right)_{023}^{\mathbf{1}}\,\mathbf{1} + \left(\mathcal{E}\right)_{023}^{\mathbf{\gamma_1}}\,\gamma_1 + \ldots \right] \nonumber \\
&+& e^0 \wedge e^3 \wedge e^1\, \left[ \, \left(\mathcal{E}\right)_{031}^{\mathbf{1}}\,\mathbf{1} + \left(\mathcal{E}\right)_{031}^{\mathbf{\gamma_1}}\,\gamma_1 + \ldots \right] \nonumber \\
&+& e^1 \wedge e^2 \wedge e^3\, \left[\, \left(\mathcal{E}\right)_{123}^{\mathbf{1}}\, \mathbf{1} + \left(\mathcal{E}\right)_{123}^{\mathbf{\gamma_1}}\,\gamma_1 + \ldots \right],
\end{eqnarray}
There are $32$ linear partial differential equations that result from \eqref{23.9}. It turns out that they split into two decoupled sectors. The first sector consists of $16$ ansatz functions: the twelve functions of 
$\vec{h}, \vec{f}, \vec{f_B}, \vec{f_C}$ and $h_{123}, f_{123}, f_{231}, f_{312}$. These $16$ ansatz functions are coupled into a set of $16$ linear partial differential equations which are given in the appendix \ref{A1}. The second decoupled sector consists of $16$ ansatz functions which are coupled into a set of $16$ linear partial differential equations which are given in the appendix \ref{A2}.

We will work with the $\gamma$-traceless gauge condition $\gamma^\mu\,\psi_\mu = 0$ which for the ansatz \eqref{21.0} results in the following equations:
\bea
\label{24.0}
{{h_{123}}} - {{f_{123}}}  - {{f_{231}}} - {{f_{312}}} = 0, \qquad  \vec{h} - \vec{f} + \vec{f_B} + \vec{f_C} = 0  \nonumber \\
-{{\vec{h}_B}} - {{\vec{f_D}}} + {{\vec{f_E}}} = 0, \qquad - {{h_0}} + {{f_{11}}}+ {{f_{22}}}+ {{f_{33}}} = 0.
\eea
Note that the first line is a condition on the ansatz functions in the first decoupled sector and the second line is a condition for the second decoupled sector. Thus $\gamma$-traceless condition is consistent with the separation of the ansatz functions into two decoupled sectors.

We now impose the gauge condition \eqref{24.0} on the equations for the first decoupled sector in \ref{A1}. We eliminate $h_{123}$ and $\vec{h}$ and obtain the following set of partial differential equations for $\vec{f}, \vec{f}_B, \vec{f}_C$, $f_{123}, f_{231}, f_{312}$. 
\be 
\label{26.0}
\vec{\nabla} \times  \left[  \frac{\vec{f_B} + \vec{f_C}}{\sqrt{V}}  \right] = 0.
\ee
\bea
\label{27.0}
\partial_x \left[ \frac{f_1-f_{C1}}{\sqrt{V}} \right] +  \partial_y \left[ \frac{f_2-f_{B2}}{\sqrt{V}} \right] = 0 \\
\label{28.0}
\partial_y \left[ \frac{f_2-f_{C2}}{\sqrt{V}} \right]  + \partial_z \left[ \frac{f_3-f_{B3}}{\sqrt{V}} \right]  = 0 \\
\label{29.0}
\partial_z \left[ \frac{f_3-f_{C3}}{\sqrt{V}} \right]  + \partial_x \left[ \frac{f_1-f_{B1}}{\sqrt{V}} \right]  = 0.
\eea
\bea
\label{30.0}
\partial_x \left[ \frac{f_3-f_{B3}}{\sqrt{V}} \right] +  \partial_y \left[ \frac{{{A_{1}}} }{\sqrt{V}} \right]   = 0 \\
\label{31.0}
\partial_y \left[ \frac{f_1-f_{B1}}{\sqrt{V}} \right]  + \partial_z \left[ \frac{{{A_{2}}} }{\sqrt{V}} \right]   = 0 \\
\label{32.0}
\partial_z \left[ \frac{f_2-f_{B2}}{\sqrt{V}} \right]  + \partial_x \left[ \frac{{{A_{3}}} }{\sqrt{V}} \right]   = 0.
\eea
\bea
\label{33.0}
\partial_y \left[ \frac{f_3-f_{C3}}{\sqrt{V}} \right] -  \partial_x \left[ \frac{{{A_{2}}} }{\sqrt{V}} \right]  = 0 \\
\label{34.0}
\partial_z \left[ \frac{f_1-f_{C1}}{\sqrt{V}} \right] - \partial_y \left[ \frac{{{A_{3}}} }{\sqrt{V}} \right]  = 0 \\
\label{35.0}
\partial_x \left[ \frac{f_2-f_{C2}}{\sqrt{V}} \right] - \partial_z \left[\frac{{{A_{1}}} }{\sqrt{V}} \right]   = 0. 
\eea
\bea
\label{36.0}
\frac{1}{\sqrt{V}} \vec{\nabla} . \left[\frac{\vec{f_B} + \vec{f_C}}{\sqrt{V}} \right] + 4 \vec{F}. (\vec{f} + \vec{f_B} + \vec{f_C}) = 0
\eea
\bea
\label{37.0}
\frac{1}{\sqrt{V}}\partial_y\left[\frac{f_3 - f_{B3}}{\sqrt{V}}\right] - \frac{1}{\sqrt{V}}\partial_z\left[\frac{f_2 - f_{C2}}{\sqrt{V}}\right] -  \frac{1}{\sqrt{V}}\partial_x\left[\frac{A_1}{\sqrt{V}}\right]   && \nonumber \\
 + 4 F_2 \left[ f_3 - f_{B3} +  f_{C3}\right]  - 4 F_3 \left[ f_2 + f_{B2} -  f_{C2}\right]  -  4 F_{1} \left[ A_1 - f_{123}\right]  &&= 0  \\
 \label{38.0}
\frac{1}{\sqrt{V}}\partial_z\left[\frac{f_1 - f_{B1}}{\sqrt{V}}\right] - \frac{1}{\sqrt{V}}\partial_x\left[\frac{f_3 - f_{C3}}{\sqrt{V}}\right] -  \frac{1}{\sqrt{V}}\partial_y\left[\frac{A_2}{\sqrt{V}}\right]   && \nonumber \\
 + 4 F_3 \left[ f_1 - f_{B1} +  f_{C1}\right]  - 4 F_1 \left[ f_3 + f_{B3} -  f_{C3}\right]  -  4 F_{2} \left[ A_2 - f_{231}\right]  &&= 0  \\
 \label{39.0}
\frac{1}{\sqrt{V}}\partial_x\left[\frac{f_2 - f_{B2}}{\sqrt{V}}\right] - \frac{1}{\sqrt{V}}\partial_y\left[\frac{f_1 - f_{C1}}{\sqrt{V}}\right] -  \frac{1}{\sqrt{V}}\partial_z\left[\frac{A_3}{\sqrt{V}}\right]   && \nonumber \\
 + 4 F_1 \left[ f_2 - f_{B2} +  f_{C2}\right]  - 4 F_2 \left[ f_1 + f_{B1} -  f_{C1}\right]  -  4 F_{3} \left[ A_3 - f_{312}\right]  &&= 0  
\eea
There are sixteen (one scalar, two vectors and one second-rank tensor) equations for twelve variables. We will address the solutions to these equations in the next section. We will not write the analogous sixteen equations for the second decoupled sector here since in this paper, we will restrict ourselves to a search for extra fermion zero modes only in the first decoupled sector.  We can think of the solutions we will obtain as having zero for the values of all the ansatz functions in the second decoupled sector.

\section{Gravitini Zero Modes beyond Broken Susies}

We will study some well-known solutions before we start to solve the equations in the first decoupled sector viz. \eqref{26.0}-\eqref{39.0}. First we will consider the fermion zero mode due to broken supersymmetries $\psi_{bs}$ given in \eqref{23.0}. The fact that $\vec{f} = \vec{f}_B = \vec{f}_C$ together with the vanishing of $f_{123}, f_{231}, f_{312}$ means that $\psi_{bs}$ solves equations \eqref{27.0}-\eqref{35.0}. Those facts together with the fact that the vectors $\vec{f}, \vec{f}_B, \vec{f}_C$ are proportional to the gradient $\vec{\nabla}V$ means that $\psi_{bs}$ also solves \eqref{26.0} and \eqref{37.0}-\eqref{39.0}. Finally we are left with \eqref{36.0} which can be directly verified. This check that $\psi_{bs}$ solves the equations we have derived  constitutes a  check on the validity of the equations.

Now we will solve the equations \eqref{26.0}-\eqref{39.0}. These are a system of coupled first order partial differential equations, to be solved on $\mathbf{R}^3$. Using \eqref{26.0}, we can write the vector field there in terms of a function $\mathcal{G}(x,y,z)$:
\be
\label{40.0}
\vec{f_B} + \vec{f_C} = \sqrt{V}~\vec{\nabla} \mathcal{G}.
\ee
Then, we consider the sum of the equations \eqref{27.0}, \eqref{28.0}, \eqref{29.0}
\be
\label{41.0}
\vec{\nabla}. \left[\frac{2\vec{f}- \vec{f_B} - \vec{f_C}}{\sqrt{V}}\right] = 0,
\ee
which together with \eqref{40.0} allows us to write the vector field $\vec{f}$ 
\be
\label{42.0}
\frac{\vec{f}}{{\sqrt{V}}} = \frac{\vec{\nabla} \mathcal{G}}{2} + \frac{\vec{\mathcal{A}}}{2}
\ee
where $\vec{\mathcal{A}}$ is a divergence-free vector field in $\mathbf{R}^3$. We have thus far used a subset of the equations to parametrise the vector fields $\vec{f}$ and $\vec{f_B}+\vec{f_C}$ in terms of a scalar $\mathcal{G}$ and a divergence-free vector field $\vec{\mathcal{A}}$. These vector fields are also the only ones which appear in the scalar equation \eqref{36.0}, which after using \eqref{40.0}, \eqref{42.0}, becomes
\be
\label{43.0}
\nabla^2 \mathcal{G} + 3 \frac{\vec{\nabla} V}{V}.\,\vec{\nabla} \mathcal{G} + \frac{\vec{\nabla} V}{V}.\,\vec{\mathcal{A}} = 0
\ee
At this stage, we note that for $\psi_{bs}$ \eqref{23.0}, $\mathcal{G}_{bs} = \frac{1}{V^2}$ and $\vec{\mathcal{A}}_{bs} = 0$. This suggests the following way of solving  \eqref{43.0}:
\be
\label{44.0}
\mathcal{G} = - \frac{C}{V^2}, \qquad   \vec{\nabla} V \, . \, \vec{\mathcal{A}} = 0,
\ee
where $\vec{\mathcal{A}}$ is now a \emph{non-zero} divergence-free vector field, $C$ is an integration constant. This choice for $\mathcal{G}$ ensures that the first summand that contributes to $\vec{f}$ in \eqref{42.0} has the right regularity properties, both at asymptotic infinity and at any of the horizons.  We now need to solve for $\vec{\mathcal{A}}$ with the right regularity properties.

Using the fact that a divergence-free vector field in three dimensions can be written as a vector cross product of two gradient vector fields \cite{Barbarosie}, together with the restriction on the divergence-free vector field coming from \eqref{44.0}, we can take 
\be
\label{45.0}
\vec{\mathcal{A}} = \vec{\nabla} \left[-\frac{1}{V^2} \right] \, \times \, \vec{\nabla} \mathcal{H}. 
\ee
where $\mathcal{H}$ is a function.  We have chosen one of the gradient vector fields to be $\vec{\nabla} \left[-\frac{1}{V^2} \right]$. This solves \eqref{44.0} and already has the required regularity properties at asymptotic infinity as well as at the horizons which means that $\mathcal{H}$ can be any function such that $\vec{\nabla} \mathcal{H}$ is $\mathcal{O}(1)$ at asymptotic infinity and is regular at any of the horizons.  We will start with the following choice 
\be
\label{46.0}
\mathcal{H}(x, y, z) = D_1\, x 
\ee
where $D_1$ is an integration constant. We now have 
\be
\label{47.0}
\vec{f} = \,C \,\left[\frac{\vec{\nabla}V}{V^{\frac52}}\right] +  D_1\,\left[ \frac{\vec{\nabla}V}{V^{\frac52}} \,\times \, (1,0,0) \right].
\ee
We have thus far completely determined $\vec{f}$ and the chosen $\mathcal{G}$ in \eqref{44.0} determines the sum $\vec{f}_B + \vec{f}_C $ \eqref{40.0} but the difference is still undetermined, which we parametrise via the vector field $\vec{L}$ as 
\bea
\label{48.0}
\frac{\vec{f}_B}{\sqrt{V}} = \,C \,\left[\frac{\vec{\nabla}V}{V^{3}}\right] - \vec{L}, \qquad 
\frac{\vec{f}_C}{\sqrt{V}} = \,C \,\left[\frac{\vec{\nabla}V}{V^{3}}\right] + \vec{L}. 
\eea
This vector field should be required to have the correct regularity properties. Substituting \eqref{47.0}, \eqref{48.0} into the equations \eqref{27.0}-\eqref{29.0}, we obtain equations for the components of $\vec{L}$ 
\begin{eqnarray}
    -\partial_x L_1+\partial_y L_2 && = - D_1\, \partial_y\left[ \frac{\partial_z V}{V^3}\right]  \nonumber \\  -\partial_y L_2 +\partial_z L_3 && = 0\nonumber\\
  - \partial_z L_3 + \partial_xL_1&&= D_1\,\partial_z\left[\frac{\partial_y V}{V^3}\right]
\end{eqnarray}{}
and the following is one solution to these equations
\be
\label{50.0}
L_1 = 0, \qquad L_2 = - D_1\left[ \frac{\partial_z V}{V^3}\right] \qquad L_3 = - D_1\left[ \frac{\partial_y V}{V^3}\right].
\ee
Thus far, we have managed to obtain the ansatz functions in $\vec{f}, \vec{f}_B, \vec{f}_C$.  We have not yet solved for the ansatz functions $f_{123}, f_{231}, f_{312}$. When we substitute the results \eqref{47.0}, \eqref{48.0}, \eqref{50.0} into the yet-to-be-solved equations \eqref{30.0}-\eqref{35.0}, we obtain equations which can be readily solved 
\be
\label{51.0}
A_1 = 2 D_1\, \frac{\partial_x V}{V^{\frac52}}, \qquad A_2 = 0, \qquad A_3 = 0
\ee 
which gives
\be
\label{52.0}
f_{123} = - D_1\, \frac{\partial_x V}{V^{\frac52}}, \qquad f_{231} = D_1\, \frac{\partial_x V}{V^{\frac52}}, \qquad f_{312} = D_1\, \frac{\partial_x V}{V^{\frac52}}.
\ee 
We have obtained all the ansatz functions in this sector, but we have not used the equations \eqref{37.0},\eqref{38.0}, \eqref{39.0}. It turns out that our solution solves these equations also. Hence we have the solution 
\bea
\label{53.0}
\vec{f} &&= \,\left( C \,\frac{\partial_xV}{V^{\frac52}}, C \,\frac{\partial_yV}{V^{\frac52}} + D_1 \,\frac{\partial_zV}{V^{\frac52}}, C \,\frac{\partial_zV}{V^{\frac52}} - D_1 \,\frac{\partial_yV}{V^{\frac52}}\right) \nonumber \\
\vec{f}_B &&= \,\left( C \,\frac{\partial_xV}{V^{\frac52}}, C \,\frac{\partial_yV}{V^{\frac52}} + D_1 \,\frac{\partial_zV}{V^{\frac52}}, C \,\frac{\partial_zV}{V^{\frac52}} + D_1 \,\frac{\partial_yV}{V^{\frac52}}\right) \nonumber \\
\vec{f}_C &&= \,\left( C \,\frac{\partial_xV}{V^{\frac52}}, C \,\frac{\partial_yV}{V^{\frac52}} - D_1 \,\frac{\partial_zV}{V^{\frac52}}, C \,\frac{\partial_zV}{V^{\frac52}} - D_1 \,\frac{\partial_yV}{V^{\frac52}}\right) \nonumber \\
f_{123} &&= - D_1\, \frac{\partial_x V}{V^{\frac52}}, \qquad f_{231} = D_1\, \frac{\partial_x V}{V^{\frac52}}, \qquad f_{312} = D_1\, \frac{\partial_x V}{V^{\frac52}}.
\eea
We can quickly generalise this solution as follows. A more general choice for $\mathcal{H}$ that goes beyond \eqref{46.0} but satisfies the required regularity conditions at asymptotic infinity and at the horizons is 
\be
\label{54.0}
\mathcal{H} = D_1\,x + D_2\,y + D_3\,z
\ee
which leads to 
\be
\vec{\mathcal{A}} = \vec{\nabla} \left[-\frac{1}{V^2} \right] \, \times \, \vec{D} ~~\text{where}~~ \vec{D} \equiv (D_1, D_2, D_3).
\ee
One can repeat along the lines of the above analysis and obtain the more general solution
\bea
\label{56.0}
\vec{f} &&= C \, \frac{\vec{\nabla} V}{V^{\frac52}} ~ + ~   \frac{\vec{\nabla} V}{V^{\frac52}}  \times \vec{D} \nonumber \\   
\vec{f}_B &&= C \, \frac{\vec{\nabla} V}{V^{\frac52}} + \left(D_2 \,\frac{\partial_zV}{V^{\frac52}} + D_3 \,\frac{\partial_yV}{V^{\frac52}} , D_3 \,\frac{\partial_xV}{V^{\frac52}} + D_1 \,\frac{\partial_zV}{V^{\frac52}}, D_1 \,\frac{\partial_yV}{V^{\frac52}} + D_2 \,\frac{\partial_xV}{V^{\frac52}} \right) \nonumber \\
\vec{f}_C &&= C \, \frac{\vec{\nabla} V}{V^{\frac52}} - \left(D_2 \,\frac{\partial_zV}{V^{\frac52}} + D_3 \,\frac{\partial_yV}{V^{\frac52}} , D_3 \,\frac{\partial_xV}{V^{\frac52}} + D_1 \,\frac{\partial_zV}{V^{\frac52}}, D_1 \,\frac{\partial_yV}{V^{\frac52}} + D_2 \,\frac{\partial_xV}{V^{\frac52}} \right) \nonumber \\
f_{123} &&= - D_1\, \frac{\partial_x V}{V^{\frac52}} + D_2\, \frac{\partial_y V}{V^{\frac52}} + D_3\, \frac{\partial_z V}{V^{\frac52}} \nonumber \\
f_{231} &&=  D_1\, \frac{\partial_x V}{V^{\frac52}} - D_2\, \frac{\partial_y V}{V^{\frac52}} + D_3\, \frac{\partial_z V}{V^{\frac52}} \nonumber \\
f_{312} &&=  D_1\, \frac{\partial_x V}{V^{\frac52}} + D_2\, \frac{\partial_y V}{V^{\frac52}} - D_3\, \frac{\partial_z V}{V^{\frac52}} \nonumber \\
\eea
\subsubsection{ The Supercharge}
The supercharge of a solution is given \cite{Aichelburg:1986wv} by 
\be
Q = -\frac{i}{k}\oint_{S^2_\infty}\gamma_5\,\gamma\wedge\psi
\ee
which for the solution in \eqref{56.0} computes to 
\begin{equation}
Q = 2 i k M \, (\gamma^i{D_i} \gamma_5 - C\, \gamma_0)\epsilon_0.
\end{equation}
The integration constant $C$ is associated with the solution from broken supersymmetries. While the solutions accompanying the three integration constants $D_1, D_2, D_3,$ are new solutions to the fermionic equations of motion. We have thus obtained extra fermion zero mode solutions. If the fermion zero modes due to broken supersymmetries is counted as one Dirac spinor worth of solutions, we have now three spinors worth of extra fermion zero modes. Since the integration constants $D_1, D_2, D_3,$ appear in the supercharge, they are physical and not some gauge artificacts.

\section{Conclusion and Future Directions}

In this paper we have provided a set up to study the solutions to the fermionic equations of motion in the background of purely bosonic supersmmetric solutions. We have done this for the Rarita-Schwinger equations for the gravitini of four dimensional $\mathcal{N}=2$ supergravity and for the Majumdar-Papapetrou solutions. We have obtained solutions to those equations beyond the ones corresponding to the broken supersymmetries. We have thus obtained some extra fermion zero mode solutions.  Clearly, there are more of these extra fermion zero modes, which is a natural study to take up. One should include the equations coming from the second decoupled sector \ref{A2}. We hope to make progress on this front \cite{work}.

The methods and techniques of this paper can be applied for other purely bosonic supersymmetric solutions of four dimensional $\mathcal{N}=2$ supergravity such as the most general solutions of \cite{Tod:1983pm}. Our goal \cite{work} is to extend our work (i) first for the Denef-Bates multi-centre solutions \cite{Bates:2003vx} in the context of four dimensional $\mathcal{N}=2$ supergravity theory that includes vector multiplets  and eventually (ii) to multi-centre solutions in the context of four dimensional $\mathcal{N}=4$ supergravity.

The structure of five dimensional $\mathcal{N}=2$ pure supergravity is very similar to it's four dimensional analog. The field content is also similar, one-forms valued in functions and spnors. A complete classification of purely bosonic solutions to this theory already exists \cite{Gauntlett:2002nw}. Very interesting bosonic solutions exist, such as a BMPV-like rotating black hole and the supersymmetric black ring \cite{Elvang:2004rt}. One hopes to develop a theory of all fermion zero mode solutions to minimal supergravity in five dimensions. 

We also need to explore the consequences of our solutions \eqref{56.0} to black hole physics. Questions such as the relation between the fermion zero modes  to the two black hole solution and the fermion zero modes of the two individual single black holes need to be answered, hopefully \cite{work}.

\begin{center}
\textbf{Acknowledgments}
\end{center}
We would like to thank Prof. Ashoke Sen for crucial discussions and guidance through out the course of this project. We would like to thank Taniya Mandal for discussions and collaboration at the early stages of the project. CNG is very thankful to the High-Energy-Section, ICTP, Trieste and in particular Prof. Bobby Acharya and Prof. Atish Dabholkar, for a visit to the ICTP in whose stimulating atmosphere a good part of this work was done.

\appendix

\section{Fermion equations}
Let $\mathcal{FE}$ denote the Clifford-algbra valued three-form equation \eqref{23.9} and let $\vec{F} =  \frac{\vec{\nabla} V}{2 V^2}$.

\subsection{\label{A1}The first decoupled sector}
The three equations $\mathcal{FE}_{012}^{~\mathbf{1}}$, $\mathcal{FE}_{023}^{~\mathbf{1}}$ and $\mathcal{FE}_{031}^{~\mathbf{1}}$ can be combined nicely into the following vector equation:
\be 
\label{a1}
\vec{\nabla} \times  \left[  \frac{\vec{h} - \vec{f}}{V^{\frac{1}{2}}}  \right] = 0
\ee
The nine equations $\mathcal{FE}_{012}^{\gamma_1\gamma_2}$,  $\mathcal{FE}_{012}^{\gamma_2\gamma_3}$, $\mathcal{FE}_{012}^{\gamma_3\gamma_1}$, $\mathcal{FE}_{023}^{\gamma_1\gamma_2}\ldots$ $\mathcal{FE}_{031}^{\gamma_3\gamma_1}$ form a rank two Cartesian tensor worth of equations:
\bea
\partial_x \left[ \frac{h_1+f_{B1}}{\sqrt{V}} \right] +  \partial_y \left[ \frac{h_2+f_{C2}}{\sqrt{V}} \right] = 0 \\
\partial_y \left[ \frac{h_2+f_{B2}}{\sqrt{V}} \right]  + \partial_z \left[ \frac{h_3+f_{C3}}{\sqrt{V}} \right]  = 0 \\
\partial_z \left[ \frac{h_3+f_{B3}}{\sqrt{V}} \right]  + \partial_x \left[ \frac{h_1+f_{C1}}{\sqrt{V}} \right]  = 0.
\eea
\bea
\partial_x \left[ \frac{h_3+f_{C3}}{\sqrt{V}} \right] +  \partial_y \left[ \frac{{{h_{123}}} - {{f_{123}}}}{\sqrt{V}} \right] = 0 \\
\partial_y \left[ \frac{h_1+f_{C1}}{\sqrt{V}} \right] + \partial_z \left[ \frac{{{h_{123}}} - {{f_{231}}}}{\sqrt{V}} \right]  = 0 \\
\partial_z \left[ \frac{h_2+f_{C2}}{\sqrt{V}} \right]  + \partial_x \left[ \frac{{{h_{123}}} - {{f_{312}}}}{\sqrt{V}} \right]  = 0.
\eea
\bea
\partial_y \left[ \frac{h_3+f_{B3}}{\sqrt{V}} \right] -  \partial_x \left[ \frac{{{h_{123}}} - {{f_{231}}}}{\sqrt{V}} \right] = 0 \\
\partial_z \left[ \frac{h_1+f_{B1}}{\sqrt{V}} \right] - \partial_y \left[ \frac{{{h_{123}}} - {{f_{312}}}}{\sqrt{V}} \right]  = 0 \\
\partial_x \left[ \frac{h_2+f_{B2}}{\sqrt{V}} \right]  - \partial_z \left[ \frac{{{h_{123}}} - {{f_{123}}}}{\sqrt{V}} \right]  = 0.
\eea
The equation $\mathcal{FE}_{123}^{\gamma_1\gamma_2\gamma_3}$ is the following scalar equation:
\bea
\frac{1}{\sqrt{V}} \vec{\nabla} . \left[\frac{\vec{f_B} + \vec{f_C}}{\sqrt{V}} \right] + 4 \vec{F}. (\vec{f} + \vec{f_B} + \vec{f_C}) = 0.
\eea
The three equations $\mathcal{FE}_{123}^{~\gamma_1}$, $\mathcal{FE}_{123}^{~\gamma_2}$, $\mathcal{FE}_{123}^{~\gamma_3}$ are the following vector equation:
\bea
\frac{1}{\sqrt{V}}\partial_y\left[\frac{f_3 - f_{B3}}{\sqrt{V}}\right] - \frac{1}{\sqrt{V}}\partial_z\left[\frac{f_2 - f_{C2}}{\sqrt{V}}\right] -  \frac{1}{\sqrt{V}}\partial_x\left[\frac{A_1}{\sqrt{V}}\right]   && \nonumber \\
 + 4 F_2 \left[ f_3 - f_{B3} +  f_{C3}\right]  - 4 F_3 \left[ f_2 + f_{B2} -  f_{C2}\right]  -  4 F_{1} \left[ A_1 - f_{123}\right]  &&= 0  \\
\frac{1}{\sqrt{V}}\partial_z\left[\frac{f_1 - f_{B1}}{\sqrt{V}}\right] - \frac{1}{\sqrt{V}}\partial_x\left[\frac{f_3 - f_{C3}}{\sqrt{V}}\right] -  \frac{1}{\sqrt{V}}\partial_y\left[\frac{A_2}{\sqrt{V}}\right]   && \nonumber \\
 + 4 F_3 \left[ f_1 - f_{B1} +  f_{C1}\right]  - 4 F_1 \left[ f_3 + f_{B3} -  f_{C3}\right]  -  4 F_{2} \left[ A_2 - f_{231}\right]  &&= 0  \\
\frac{1}{\sqrt{V}}\partial_x\left[\frac{f_2 - f_{B2}}{\sqrt{V}}\right] - \frac{1}{\sqrt{V}}\partial_y\left[\frac{f_1 - f_{C1}}{\sqrt{V}}\right] -  \frac{1}{\sqrt{V}}\partial_z\left[\frac{A_3}{\sqrt{V}}\right]   && \nonumber \\
 + 4 F_1 \left[ f_2 - f_{B2} +  f_{C2}\right]  - 4 F_2 \left[ f_1 + f_{B1} -  f_{C1}\right]  -  4 F_{3} \left[ A_3 - f_{312}\right]  &&= 0  
\eea
where we have defined
\bea
{{A_{1}}} \equiv {{f_{312}}} + {{f_{231}}}, \quad {{A_{2}}} \equiv {{f_{123}}} + {{f_{312}}} \quad {{A_{3}}} \equiv {{f_{231}}} + {{f_{123}}}.
\eea

\subsection{\label{A2}The second decoupled sector}
The equation $\mathcal{FE}_{123}^{~\mathbf{1}}$ is the following scalar equation:
\bea
{{\vec{\nabla}.\vec{f}_D}} + \frac32  \frac{\vec{\nabla}V .  {{\vec{f}_D}}}{V}  = 0.
\eea
The nine equations $\mathcal{FE}_{012}^{\gamma_1}$,  $\mathcal{FE}_{012}^{\gamma_2}$, $\mathcal{FE}_{012}^{\gamma_3}$, $\mathcal{FE}_{023}^{\gamma_1}\ldots$ $\mathcal{FE}_{031}^{\gamma_3}$ form a rank two Cartesian tensor worth of equations:
\bea
  \frac{1}{\sqrt{V}} \, \partial_x \left[ \frac{{{h_{23}}} + {{f_{23}}}}{\sqrt{V}} \right] + \frac{1}{\sqrt{V}} \, \partial_y \left[ \frac{{{h_{31}}} - {{f_{13}}}}{\sqrt{V}} \right]  +  4 F_{1} {{f_{23}}} - 4 F_2  {{f_{13}}}  + 4F_{3}{{h_{12}}} && = 0\\
  \frac{1}{\sqrt{V}} \, \partial_y \left[ \frac{{{h_{31}}} + {{f_{31}}}}{\sqrt{V}} \right] + \frac{1}{\sqrt{V}} \, \partial_z \left[ \frac{{{h_{12}}} - {{f_{21}}}}{\sqrt{V}} \right]  +  4 F_{2} {{f_{31}}} - 4 F_3  {{f_{21}}}  + 4F_{1}{{h_{23}}} && = 0\\
  \frac{1}{\sqrt{V}} \, \partial_z \left[ \frac{{{h_{12}}} + {{f_{12}}}}{\sqrt{V}} \right] + \frac{1}{\sqrt{V}} \, \partial_x \left[ \frac{{{h_{23}}} - {{f_{32}}}}{\sqrt{V}} \right]  +  4 F_{3} {{f_{12}}} - 4 F_1  {{f_{32}}}  + 4F_{2}{{h_{31}}} && = 0
\eea

\bea
 \frac{1}{\sqrt{V}} \, \partial_x \left[ \frac{{{h_{0}}} - {{f_{22}}}}{\sqrt{V}} \right] 
 + \frac{1}{\sqrt{V}} \, \partial_y \left[ \frac{{{h_{12}}} + {{f_{12}}}}{\sqrt{V}} \right]
- 4 F_{1} {{f_{22}}}  + 4 F_2 {{f_{12}}}  - 4F_{3}{{h_{31}}} && = 0 \\
 \frac{1}{\sqrt{V}} \, \partial_y \left[ \frac{{{h_{0}}} - {{f_{33}}}}{\sqrt{V}} \right] 
 + \frac{1}{\sqrt{V}} \, \partial_z \left[ \frac{{{h_{23}}} + {{f_{23}}}}{\sqrt{V}} \right]
- 4 F_{2} {{f_{33}}}  + 4 F_3 {{f_{23}}}  - 4F_{1}{{h_{12}}} && = 0 \\
 \frac{1}{\sqrt{V}} \, \partial_z \left[ \frac{{{h_{0}}} - {{f_{11}}}}{\sqrt{V}} \right] 
 + \frac{1}{\sqrt{V}} \, \partial_x \left[ \frac{{{h_{31}}} + {{f_{31}}}}{\sqrt{V}} \right]
- 4 F_{3} {{f_{11}}}  + 4 F_1 {{f_{31}}}  - 4F_{2}{{h_{23}}} && = 0 
\eea

\bea
\frac{1}{\sqrt{V}} \, \partial_x \left[ \frac{{{h_{12}}} - {{f_{21}}}}{\sqrt{V}} \right]  - \frac{1}{\sqrt{V}} \, \partial_y \left[ \frac{{{h_{0}}} - {{f_{11}}}}{\sqrt{V}} \right] - 4 F_{1} {{f_{21}}} + 4 F_2 {{f_{11}}}  - 4F_{3}{{h_{23}}} && = 0 \\
\frac{1}{\sqrt{V}} \, \partial_y \left[ \frac{{{h_{23}}} - {{f_{32}}}}{\sqrt{V}} \right]  - \frac{1}{\sqrt{V}} \, \partial_z \left[ \frac{{{h_{0}}} - {{f_{22}}}}{\sqrt{V}} \right] - 4 F_{2} {{f_{32}}} + 4 F_3 {{f_{22}}}  - 4F_{1}{{h_{31}}} && = 0 \\
\frac{1}{\sqrt{V}} \, \partial_z \left[ \frac{{{h_{31}}} - {{f_{13}}}}{\sqrt{V}} \right]  - \frac{1}{\sqrt{V}} \, \partial_x \left[ \frac{{{h_{0}}} - {{f_{33}}}}{\sqrt{V}} \right] - 4 F_{3} {{f_{13}}} + 4 F_1 {{f_{33}}}  - 4F_{2}{{h_{12}}} && = 0 
\eea

The three equations $\mathcal{FE}_{123}^{\gamma_1\gamma_2}$, $\mathcal{FE}_{123}^{\gamma_2\gamma_3}$, $\mathcal{FE}_{123}^{\gamma_3\gamma_1}$ are:
\bea
\frac{1}{\sqrt{V}} \, \partial_x \left[ \frac{{{f_{31}}} -  {{f_{2123}}}}{\sqrt{V}}  \right] + \frac{1}{\sqrt{V}} \, \partial_y \left[ \frac{{{f_{32}}} +  {{f_{1123}}}}{\sqrt{V}}  \right]  -  \frac{1}{\sqrt{V}} \, \partial_z \left[ \frac{{{f_{11}}} +  {{f_{22}}}}{\sqrt{V}} \right]  \nonumber \\
+  4 F_1 ({{f_{31}}} -  {{f_{2123}}}) - 4 F_{2} ({{f_{32}}} +  {{f_{1123}}}) && = 0 \\
\frac{1}{\sqrt{V}} \, \partial_y \left[ \frac{{{f_{12}}} -  {{f_{3123}}}}{\sqrt{V}}  \right] + \frac{1}{\sqrt{V}} \, \partial_z \left[ \frac{{{f_{13}}} +  {{f_{2123}}}}{\sqrt{V}}  \right]  -  \frac{1}{\sqrt{V}} \, \partial_x \left[ \frac{{{f_{22}}} +  {{f_{33}}}}{\sqrt{V}} \right]  \nonumber \\
+  4 F_2 ({{f_{12}}} -  {{f_{3123}}}) - 4 F_{3} ({{f_{13}}} +  {{f_{2123}}}) && = 0 \\
\frac{1}{\sqrt{V}} \, \partial_z \left[ \frac{{{f_{23}}} -  {{f_{1123}}}}{\sqrt{V}}  \right] + \frac{1}{\sqrt{V}} \, \partial_x \left[ \frac{{{f_{21}}} +  {{f_{3123}}}}{\sqrt{V}}  \right]  -  \frac{1}{\sqrt{V}} \, \partial_y \left[ \frac{{{f_{33}}} +  {{f_{11}}}}{\sqrt{V}} \right]  \nonumber \\
+  4 F_3 ({{f_{23}}} -  {{f_{1123}}}) - 4 F_{1} ({{f_{21}}} +  {{f_{3123}}}) && = 0 
\eea
The three equations $\mathcal{FE}_{~023}^{\gamma_1\gamma_2\gamma_3},~ \mathcal{FE}_{~031}^{\gamma_1\gamma_2\gamma_3},~ \mathcal{FE}_{~012}^{\gamma_1\gamma_2\gamma_3}$ are:
\bea
\frac{1}{\sqrt{V}} \, \partial_z \left[ \frac{{{h_{31}}} - {{f_{2123}}}}{\sqrt{V}} \right]  - \frac{1}{\sqrt{V}} \, \partial_y \left[ \frac{{{h_{12}}} - {{f_{3123}}}}{\sqrt{V}} \right] &&\nonumber \\
+ F_2 ({{h_{12}}} + 3{{f_{3123}}} - {{h_0}} +{{f_{33}}} ) 
- F_3 ({{h_{31}}} + 3{{f_{2123}}} + {{h_{23}}}  + {{f_{23}}})  + 4 F_{1} {{h_{12}}} &&= 0 \\
\frac{1}{\sqrt{V}} \, \partial_x \left[ \frac{{{h_{12}}} - {{f_{3123}}}}{\sqrt{V}} \right]  - \frac{1}{\sqrt{V}} \, \partial_z \left[ \frac{{{h_{23}}} - {{f_{1123}}}}{\sqrt{V}} \right] &&\nonumber \\
+ F_3 ({{h_{23}}} + 3{{f_{1123}}} - {{h_0}} +{{f_{11}}} ) 
- F_1 ({{h_{12}}} + 3{{f_{3123}}} + {{h_{31}}}  + {{f_{31}}})  + 4 F_{2} {{h_{23}}} &&= 0 \\
\frac{1}{\sqrt{V}} \, \partial_y \left[ \frac{{{h_{23}}} - {{f_{1123}}}}{\sqrt{V}} \right]  - \frac{1}{\sqrt{V}} \, \partial_x \left[ \frac{{{h_{31}}} - {{f_{2123}}}}{\sqrt{V}} \right] &&\nonumber \\
+ F_1 ({{h_{31}}} + 3{{f_{2123}}} - {{h_0}} +{{f_{22}}} ) 
- F_2 ({{h_{23}}} + 3{{f_{1123}}} + {{h_{12}}}  + {{f_{12}}})  + 4 F_{3} {{h_{31}}} &&= 0.
\eea

\end{document}